# High-Throughput Biologically Optimized Search Engineering Approach to Synthetic Biology

A. X. C. N. Valente[1,2,3] and Stephen S. Fong[3,4]

[1]Biocant - Biotechnology Innovation Center, Cantanhede 3060-183, Portugal
[2]Center for Neuroscience and Cell Biology, University of Coimbra, Coimbra 3004-517, Portugal
[3]Center for the Study of Biological Complexity, Virginia Commonwealth University, Richmond, VA 23284-3028, USA
[4]Dept. of Chemical and Life Sciences Engineering, Virginia Commonwealth University, Richmond, VA 23284-3028, USA

**Abstract**

Synthetic Biology is the new engineering-based approach to biology that includes applications of designing complex biological devices. At present, it is not yet clear what will emerge as the defining principles of Synthetic Biology. One proposed approach is to build Synthetic Biology around the classical engineering principles of standardization, modularity/decoupling and abstraction/modeling to facilitate component-based design. In this article we suggest and discuss an alternative paradigm, which we call High-throughput Biologically Optimized Search Engineering (HT-BOSE). Stemming from directed evolution, in HT-BOSE the focal point is a biological knowledge based rational optimization of the search process in the space of device design possibilities. The HT-BOSE approach may also be relevant in other contexts and we briefly highlight how it could be applicable to the development of multi-drug cocktails in a biomedical setting.

**Introduction**

Genetic engineering, defined as the purposeful manipulation of a single gene within an organism, is today a commonplace technology (1). A greater challenge is moving from single-gene manipulation, to the engineering of complex biological devices that incorporate multiple interacting functions. The goal at present is creation of a full-fledged engineering field with living matter as its substrate. This hopeful engineering field is most often called Synthetic Biology (2).

Every engineering field is characterized by its own set of canonical principles. These principles are distinct insofar as they are determined by different substrates of study, engineering objectives and relevant governing physical laws. Nonetheless, all major engineering fields are based on an overarching common design philosophy (2). Sustained by the three concepts of standardization of parts, modularity/decoupling and abstraction/modeling, this Classical Engineering Design (CED) approach allows the construction of complex engineering devices with manageable, limited trial and error tuning. This is remarkable, as a very large number of parts and an even larger number of interactions between those parts would normally render the successful implementation of a device all but impossible.

There is an ongoing major effort to build a CED foundation for Synthetic Biology (3). For instance, a registry of standard biological parts has been created (4). At present it contains thousands of distinct entries, with varying degrees of quality control for each part. This foundation already allows undergraduate students in a yearly summer Synthetic Biology competition to, in mere weeks, successfully build working devices of a few parts (5). However, the ultimate goal of Synthetic Biology is not being able to put together elementary devices in days or weeks, but rather to enable the engineering of complex devices.

In Synthetic Biology unwanted interactions are inherently difficult to predict and avoid (6). This is due to the molecular/chemical nature of the interactions and the unfeasibility of physically compartmentalizing every interaction. As the number of parts in a device grows, the number of such pernicious interactions will grow even faster. We argue this scaling poses a major hurdle for traditional CED component-based approaches and potentially renders CED an ineffective approach to engineer complex biological devices. For a device involving dozens of parts, the amount of fine tuning required would be unworkable.

A distinct conceptual view of engineering is attained by considering Design Space (DS) - the space of all possible device designs. Given the time to try out every possible design, it would be a mere formality to select the best design to the task at hand. Naturally, this is not feasible due to the very high dimensionality of DS. CED provides a way of immediately placing ourselves close to a working design, which can then be hopefully reached via some fine-tuning. However, as explained above, we argue that for complex biological devices, CED is not a good approach as in this case the large number of pernicious interactions will actually place us far from a functioning design.

The evolutionary histories of natural organisms trace paths in DS. This motivated the biological engineering technique of directed evolution (7). Directed evolution consists of performing alternating rounds of mutation and selection/screening with the intent of moving in DS towards a good design for the task at hand. It has been extensively applied to the engineering of enzymes (8), and more recently to the engineering of elementary biological devices (9). Nevertheless, so far it has not been possible to construct complex devices via directed evolution any more than it has been via CED. Directed evolution as a nonspecific statistical search technique in a very high-dimensional space of possibilities has not proven to be powerful enough.

We suggest that the future of Synthetic Biology lies in combining biological knowledge and directed evolution into a biologically-guided rational optimization of the DS search process. We designate this approach High-Throughput Biologically Optimized Search Engineering (HT-BOSE) and suggest it as a framework for designing biological devices. This article will focus on founding principles for HT-BOSE. In parallel, HT-BOSE will also require the development of appropriate technologies for its implementation, in particular for an increase in the throughput of the search process (10; 11; 12). We consider there are three foundational elements to an HT-BOSE approach, which we now discuss in turn: Search strategies, the Design Registry and fitness assays.

**Search strategies**

We start by posing a streamlined version of the search problem in HT-BOSE. Given a particular aim for a device, every design in DS has an associated fitness score. The fitness of any given design is unknown, until the design is realized and assayed in some fashion. The objective is to find a high fitness design, under the constraint that a limited number of designs can be assayed.

If there is no knowledge at all of the structure of the fitness function over DS, then the strategy of randomly picking designs in DS for assaying is as good as any other search strategy. This is an instance of the principle that "there are no free-lunches in optimization" (13).

Now, it is clear that designs that differ by a single base pair in their encoding DNA tend to have correlated fitness, as they tend to be designs with modest differences. This *DS functional structure* enables the pursuit of the traditional hill-climbing search, i.e. directed evolution, based on point mutations (7). Namely, a search that is based on successive rounds of point mutation and selection of the best mutated design(s) as the starting point(s) for the next round. Naturally, this hill climbing towards a higher fitness design is a much more efficient strategy than random search. Nonetheless, mutation interactions can be nonlinear with regards to effect on fitness. For instance, a simultaneous mutation in multiple sites might be beneficial, whereas the same mutations in isolation are not. As a result, many design improvements are unreachable via point mutation hill-climbing. In other words, point mutation hill-climbing is limited by local fitness optima.

Yet, there is a much richer DS functional structure than the one just described. For instance, there is DS functional structure at other length-scales. Knowledge of this additional biological structure can be explored to improve the hill-climbing search process. We start by describing optimization of the search process at a very different length-scale.

Microbial consortia

Community level synergies between different microbial species often result in a consortium of microbes being more efficient than a single organism at fulfilling a desired task (14). As an example, microbial communities might be pertinent to biofuel production (15). In the context of consortia, a point in DS represents a potential Community Design (defined by a design being ascribed to each organism in the community). A community can be optimized by the point mutation search process previously described, with the point mutations being applied over all the microbes in the community. However, there is also DS functional structure at the level of what microbes constitute the community. Therefore, an additional way of hill-climbing in DS is via whole-microbe-at-a-time changes to the community composition. Note how this constitutes searching DS via jumps at a distinct length-scale that is much larger than point mutation changes. In a sense, swapping microbes can be viewed as a course-grained adjustment within DS, whereas point mutations can be viewed as a fine-grained adjustment.

Given the biological significance of genes, there is invariably DS functional structure at the level of what genes are present in an organism. Hill-climbing in DS at the gene length-scale, via whole-gene-at-a-time changes to an organism, is therefore also pertinent. Gene length-scale changes would stand as an intermediate-grained adjustment compared to whole organism exchanges (course-grained) and genetic point mutations (fine-grained).

Recombination based techniques are widely used in the enzyme optimization context (16; 17). For this family of techniques, the hill-climbing in DS occurs via recombination of entire parts of proteins. The effectiveness of the method rests on the modular nature of proteins, that constitutes DS functional structure at yet another length-scale (18).

The examples above leverage generic DS functional structure. However, biological knowledge specific to the problem at hand can be invaluable. High-throughput genomic data collection and analysis can play an important role in this regard (19; 20; 21). As a case in point, many of these techniques yield functional networks of genes. Such networks embody DS functional structure information. Analogously to what happens with mutations, gene interactions can have a nonlinear effect on fitness. The simultaneous introduction of multiple genes in an organism might be beneficial, while the isolated introduction of any one of those same genes is not. Functional networks of genes enable hill-climbing strategies based on prioritizing specific multiple-genes-at-a-time changes to an organism. An otherwise unguided search at a multiple-genes-at-a-time-jumps length-scale would be generally unsuccessful, due to the volume of DS to be explored and the relative sparseness of worthy designs in DS. Functional networks at other biological levels may likewise be explored.

In CED, knowledge is often applied directly to the design of a device. It is not easy to acquire biological knowledge at the level of detail required for relevance in such a task. In HT-BOSE, biological knowledge is equally important. However, as shown, it is instead used to infer DS functional structure that enables a more efficient search process. We suggest this less demanding form of leveraging knowledge is more realistic for the complex biological setting and, in particular, may better match the typical detail-level of information generated by high-throughput approaches.

Ultimately, HT-BOSE will involve combining a diversity of hill-climbing strategies in a true multi-scale search process for engineering function. Figure 2 outlines how this could be implemented in practice. Now, a successful search in a very high-dimensional space, besides an efficient search strategy, requires an appropriate choice of where to start the search. We next discuss implications of this to an HT-BOSE approach to Synthetic Biology.

**Design Registry for HT-BOSE**

In CED, engineering of a new device relies on combining and reusing existing parts. A good library of standardized parts, appropriately catalogued and characterized, is crucial to enable this. For engineering of devices via a search process, the counterpart is a good library of *Seed Designs* to serve as search starting points. Further, such a Design Registry should also be a source of the seed-parts required for the ensuing jumps-at-different-length-scales search strategies discussed in the previous section.

A seed likely close in DS to a design that fulfills the engineering objectives at hand has obvious advantages. An existing microbe that already nearly fulfills the new engineering objectives constitutes naturally a good starting point for modifications. However, the choice of ideal starting point is not always so clear. There may be no existing device already very close to fulfilling the desired engineering objectives. For example, the objectives may require a combination of traits of vastly different existing organisms. An important characteristic for a seed is that of flexibility. Some designs may not be practical engineering devices for any purpose and yet constitute excellent seeds. That is the case if the structure of DS is such that it is possible to hill-climb from them to a large diversity of useful designs. Again, from the versatile-community to the versatile-single gene, such flexible designs and design-parts are present at all levels.

In contrast to the versatile design, a design may be an excellent practical device for a given purpose and yet, live in a region of DS that makes it a not so good starting point for hill-climbing to a design that addresses any objective other than that one. A trade-off between specificity (in the sense of closeness to fulfilling the final desired objective) and versatility may have to be considered when choosing seeds for a search process. The successful selection of seeds will not be trivial and likely will require trial and error. It will constitute a major part of the process of engineering a new device via HT-BOSE.

The seed selection process will rest on information available in the Design Registry. A design entry in the Registry must, besides data on functional characteristics, record how that design was generated. When an existing design is used as a seed for subsequent designs, this must also be noted. Effectively, designs can have a forward phylogenetic history in addition to their backward phylogenetic history. Sliding along a phylogeny may allow in particular an adjustment of the specificity-flexibility trade-off in seed selection.

The complexity of the phylogenetic history of a design is compounded by the fact that the search strategies discussed do not necessarily lead to elementary relationships between designs. The search at different length-scales process creates devices that have received contributions from a multiplicity of sources, across a range of length-scales. The discovery of horizontal gene transfer has complicated natural phylogenetics, but a phylogenetics for designs resulting from HT-BOSE is likely to be even more complex. Developing an ontology that facilitates handling of all this information will be of essence, given the crucial role of seed selection in the HT-BOSE engineering approach.

In its early stages, a Design Registry for HT-BOSE should take maximum advantage of the enormous diversity of communities, microbes and genes that Nature already provides. In this regard, it stands to benefit greatly from large-scale metagenomic data collection efforts (22; 23). As new synthetic designs are engineered and added to the Design Registry, older entries gradually develop forward phylogenies. In due course, the more flexible designs and design-parts become apparent, making the Design Registry ever more valuable in facilitating the seed selection phase of HT-BOSE.

We have discussed seed designs, where DS search processes begin. We have also already addressed search strategies for generating designs to assay in subsequent rounds of a search process. It remains to focus on the third major element of an HT-BOSE approach to Synthetic Biology, the fitness scoring assay.

**Fitness assays**

Fitness assaying is the process whereby select designs are assayed and assigned a fitness score. For HT-BOSE, the assaying and fitness scoring should ideally be amenable to high-throughput implementation. The best designs, based on their scores, are then utilized by the search strategy to generate the next set of designs to fitness assay. Iteration of this procedure hopefully leads to hill-climbing in DS towards a design that accomplishes the engineering objective at hand.

The natural fitness assay to use is the one that assesses how well a design fulfills the desired engineering objective. Unfortunately, this straightforward assay choice may not necessarily succeed. When the designs being assayed are still far from accomplishing the desired engineering purpose, a fitness score based on proximity to satisfying that goal may lack the *scoring resolution* for differentiating between those designs. The designs may all appear equally bad at achieving the objective, therefore precluding hill-climbing via prioritization of the best ones. In such situations, the HT-BOSE approach requires devising a succession of *transitional fitness assays*. These transitional fitness assays must provide intermediate goals that are interspersed closely enough for adequate fitness scoring resolution to exist at all times. Biological insight must be used to ensure that fulfillment of the succeeding intermediate goals corresponds to getting progressively closer to a design that meets the final engineering goal. Successful transitional fitness assays are likely rather problem-specific and therefore their formulation constitutes one of the more challenging tasks in HT-BOSE.

A switch to hill-climbing under an alternative *supporting fitness assay* may be valuable in instances the search in DS gets trapped in a local optima with respect to the main (and ultimately relevant) fitness function. Again, the choice of a supporting fitness assay should rely on biological insight into the specific problem in question. An intelligent choice for the supporting fitness assay could greatly surpass the typical default option of random jumping in DS every time trapping in a local optima occurs. A particularly relevant characteristic for a supporting assay may be the ability to draw the search towards a more versatile region in DS, before hill-climbing of the main fitness function towards the ultimate goal is resumed. Note how the specificity-flexibility adjustment, previously mentioned in the context of seed design selection, is now occurring during the actual search, via alternating between the main and the supporting fitness assay. A case in point is the use of a thermostability assay as the supporting assay in enzyme optimization problems (24).

Dynamically, optimization under a single fitness function can be visualized as a straightforward monotonic hill-climb towards a local fitness peak. By comparison, optimization under alternating fitness functions can produce a much richer spectrum of dynamical behaviors. As the simplest example, periodic orbits are possible in this latter case. It will be interesting to further explore from a mathematical standpoint this class of switched dynamical systems.

We end this section with reminders related to two peculiarities of engineering in a biological context. First, synthetic biological devices must be engineered for evolutionary stability under the setting and time-scale on which they will function. Therefore the fitness assay must be formulated to properly assess the operationally relevant evolutionary end point of a design. Second, a fundamental premise for pursuing HT-BOSE is that unwanted biological interactions are inherently difficult to circumvent. Therefore the subject of any fitness assaying should be the full integrated device at all times. Unlike under a CED approach, it makes no sense to develop different parts of a device in isolation and combine them only at a later stage.

Although we have been discussing HT-BOSE solely in the context of Synthetic Biology, it is applicable to other problems involving search within a very large space of possibilities. In the next section, we highlight the relevance of HT-BOSE in a particular biomedical setting.

**HT-BOSE in medical drug discovery**

Medical drug discovery has been dominated by the single compound, single molecular target, paradigm. However, many common diseases, such as cancer, diabetes, immune-inflammatory and cardio-vascular disorders are likely inherently system-level pathologies (25). There is a growing appreciation that trying to address such system-level disruptions via modulation of a single target is unnecessarily restrictive. Instead, the single drug magic-bullet could be beneficially replaced by the multi-compound drug cocktail, that synergistically acts at multiple points of the system (26; 27). But direct design of a drug cocktail is too difficult while, akin to the DS of Synthetic Biology, the Cocktail Space of a priori possible drug cocktails is too large for a blind exhaustive search. A credible alternative may be biologically rational optimization of the search process. The HT-BOSE approach thus becomes directly relevant to the drug cocktail discovery problem. As an elementary example, Drug Cocktail Space functional structure exists in the form of compounds that are co-present in natural organisms. Medically this is reflected in the oftentimes apparent more positive therapeutic effect of a traditional medicine natural product versus that of its purified, most significant compound in isolation (27).

**Final Remarks**

We propose that applying biological knowledge to the rational optimization of a DS search process is the most appropriate paradigm for Synthetic Biology. We discussed an elementary foundation for this HT-BOSE approach. In particular, we indicated how in HT-BOSE biological knowledge would be leveraged: via appropriate choice of seed designs, search strategies (based on DS functional structure) and transitional and supporting fitness assays. Notably, the HT-BOSE approach underscores harnessing the unforeseeable, but likely plentiful and highly valuable, synergies possible between existing biological elements.

Although we stressed contrasts between the two, there is effectively a continuum from pure HT-BOSE to pure CED. Different Synthetic Biology problems will likely require approaches at different points within this spectrum, seldom at the two extremes.

Synthetic Biology and Systems Biology, if successful, will bring with them a greatly enhanced understanding of living organisms, through the ability to engineer them and the ability to predict their behavior (28; 29). Hopefully the power of this knowledge will lead to an ever greater respect and appreciation for the preciousness of Life (30; 31).

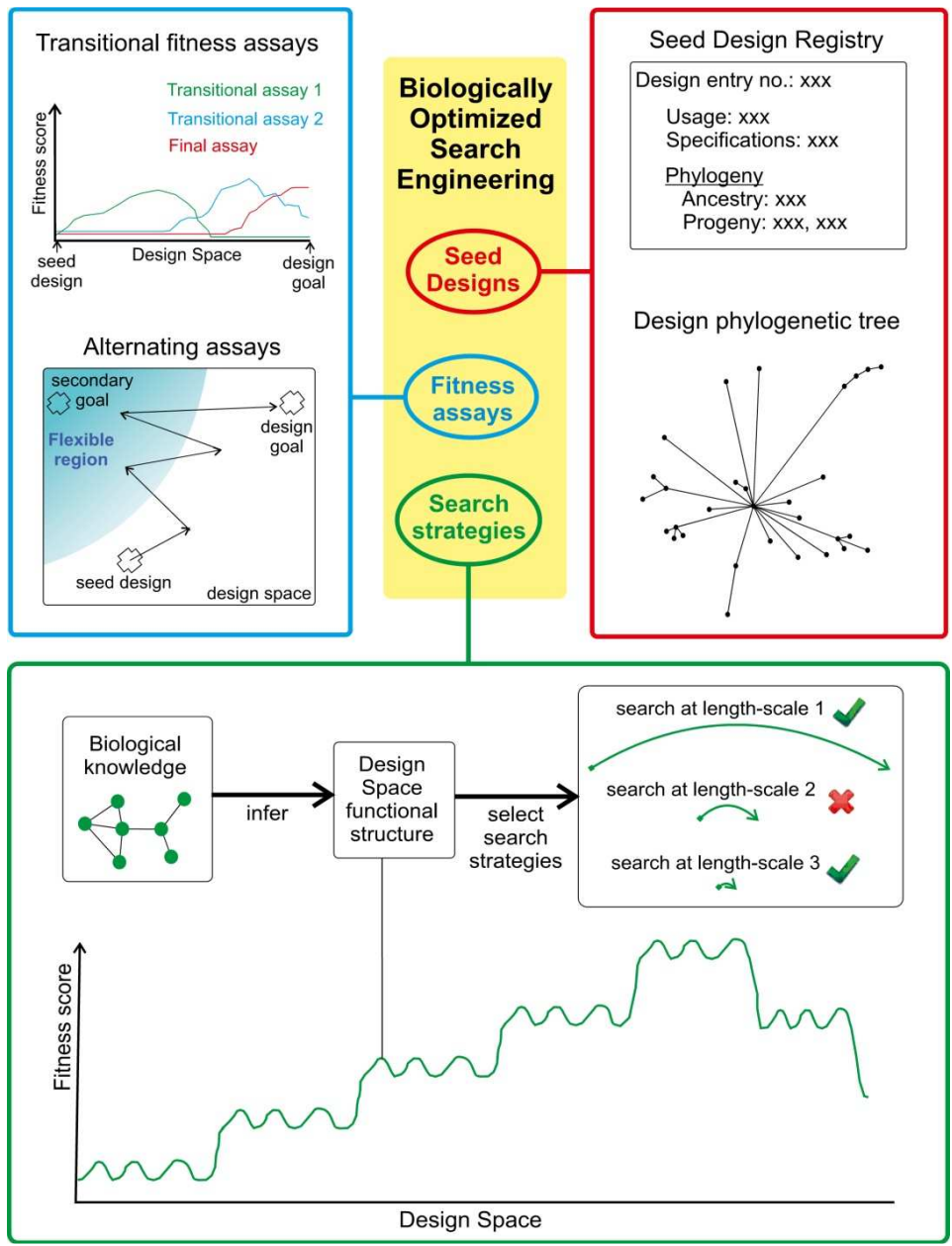

**Figure 1. The HT-Biologically Optimized Search Engineering approach to Synthetic Biology.**
**Seed Designs:** A Registry of Seed Designs provides starting points (seeds) for the search processes in Design Space. To facilitate seed selection, Registry records should include phylogenetic information on how designs were generated, as well as on what subsequent designs originated from them. Seed selection may involve considering the trade-off between closeness of a seed to fulfilling the desired engineering objective and versatility of the seed as a search starting point. Designs that show high versatility as search starting points are apparent in the design phylogenetic tree. **Fitness assays:** Fitness scores are assigned to designs tested in the fitness assay. If assayed designs are still too far from fulfilling the desired goal, an evaluation based on this fulfillment may be unable to differentiate between the designs. In such cases, a succession of transitional fitness assays associated with intermediate engineering goals must be employed. Search processes can also get trapped in local fitness optima. Alternating the main assay with a supporting assay may circumvent this problem. A supporting assay that draws the search towards a more flexible region of Design Space may prove particularly helpful. **Search strategies:** Biological data is seldom detailed enough to allow a direct design approach. Yet, it may permit inference of Design Space functional structure. Devising rational Design Space search strategies is then possible. In particular, Design Space functional structure at multiple length-scales should be explored.

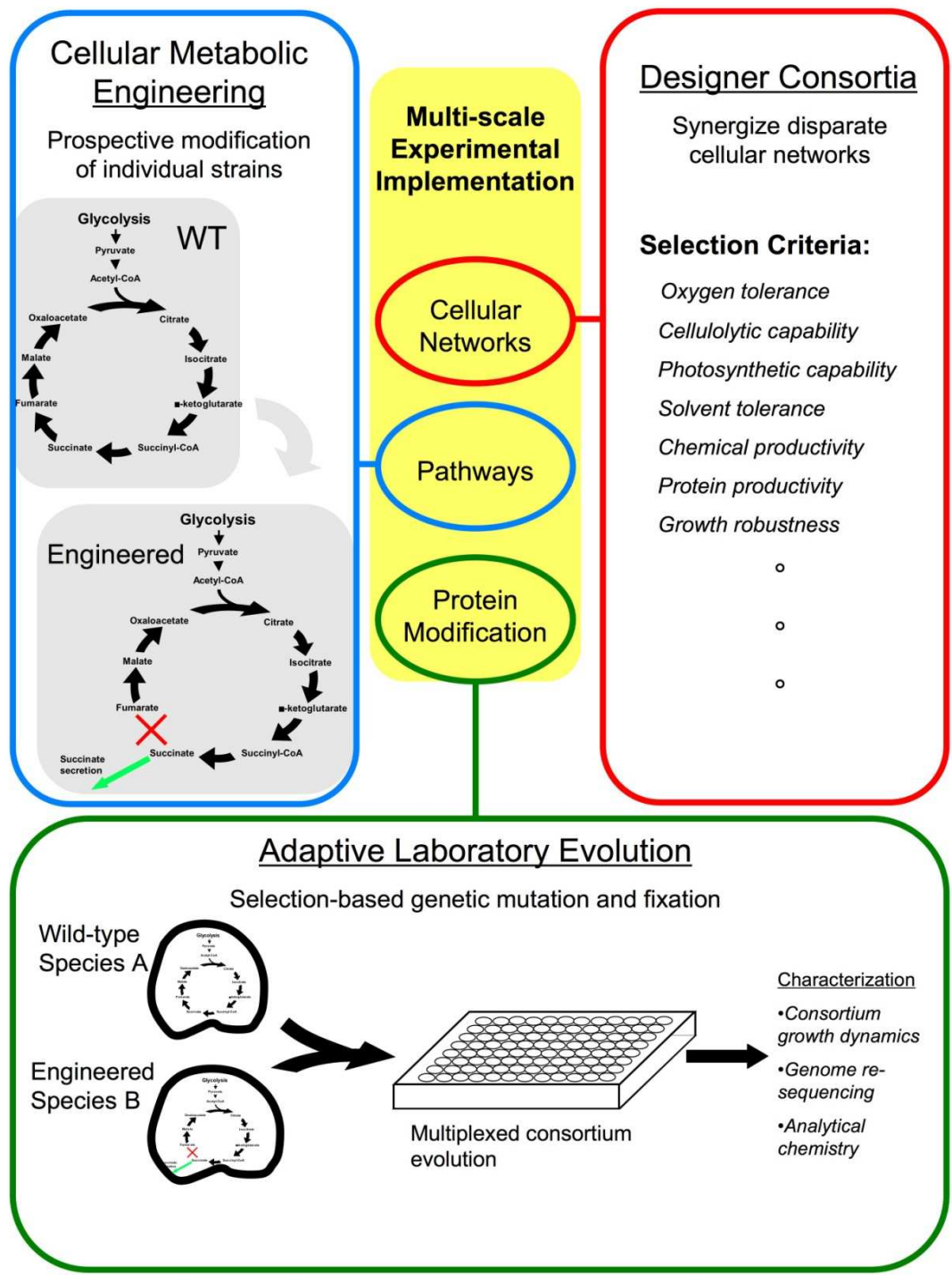

**Figure 2. Outline of a microbial consortia optimization experiment.**
Multi-scale experimental implementation of HT-BOSE will utilize cellular, pathway and molecular/genetic experimental tools to parallel computational analyses. Different organisms can be screened and selected based upon favorable functional characteristics to define the initial composition of a **Designer consortium**. Each of the selected organisms can be modified by pathway engineering techniques to delete or add biochemical capabilities (**Metabolic engineering**). Molecular and genetic changes can be implemented through extended co-culturing of selected and/or engineered organisms to foster stable consortia using **Adaptive laboratory evolution**.